\documentclass{sig-alternate}
\usepackage{url}
\begin{document}
%



\title{Disaggregated and optically interconnected memory: when will it be cost effective?
}

%
%
%
%
%

\numberofauthors{5} 

%

%
%

\author{
\alignauthor
Bulent Abali\\
       \affaddr{IBM T.J. Watson Research Center}\\
       \affaddr{Route 134}\\
       \affaddr{Yorktown Heights, NY 10598}\\
       \email{abali@us.ibm.com}
\alignauthor Richard J. Eickemeyer\\
       \affaddr{IBM STG Group}\\
       \affaddr{3605 HWY 52 N}\\
       \affaddr{Rochester MN 55901}\\
       \email{eick@us.ibm.com}
\alignauthor
Hubertus Franke\\
       \affaddr{IBM T.J. Watson Research Center}\\
       \affaddr{Route 134,} \\
       \affaddr{Yorktown Heights, NY 10598}\\
       \email{frankeh@us.ibm.com}
\and     
\alignauthor 
Chung-Sheng Li\\
       \affaddr{IBM T.J. Watson Research Center}\\
       \affaddr{Route 134,} \\
       \affaddr{Yorktown Heights, NY 10598}\\
       \email{csli@us.ibm.com}
\alignauthor Marc A. Taubenblatt\\
       \affaddr{IBM T.J. Watson Research Center}\\
       \affaddr{Route 134,} \\
       \affaddr{Yorktown Heights, NY 10598}\\
       \email{tauben@us.ibm.com}
}




\maketitle
\begin{abstract}

The ``Disaggregated Server'' concept has been proposed for datacenters
where the same type server resources are aggregated in their
respective pools, for example a compute pool, memory pool, network
pool, and a storage pool.  Each server is constructed dynamically by
allocating the right amount of resources from these pools according to
the workload's requirements.  Modularity, higher packaging and cooling
efficiencies, and higher resource utilization are among the suggested
benefits.  With the emergence of very large datacenters, ``clouds''
containing tens of thousands of servers, datacenter efficiency has
become an important topic.  Few computer chip and systems vendors
are working on and making frequent announcements on silicon photonics
and disaggregated memory systems.

In this paper we study the trade-off between cost and performance of
building a disaggregated memory system where DRAM modules in the
datacenter are pooled, for example in memory-only chassis and racks.
The compute pool and the memory pool are interconnected by an optical
interconnect to overcome the distance and bandwidth issues of
electrical fabrics.  We construct a simple cost model that includes
the cost of latency, cost of bandwidth and the savings expected from a
disaggregated memory system.  We then identify the level at which a
disaggregated memory system becomes cost competitive with a
traditional direct attached memory system.  

Our analysis shows that a rack-scale disaggregated memory system will
have a non-trivial performance penalty, and at the datacenter scale
the penalty is impractically high, and the optical interconnect costs
are at least a factor of 10 more expensive than where they should be
when compared to the traditional direct attached memory systems.

\end{abstract}



\terms{Memory, Cost, Performance, Disaggregated Servers, Silicon Photonics}


\section{Introduction}

Traditional servers suffer from a resource fragmentation problem where
one or more of the resources --compute, memory, network, or storage--
become underutilized because of mismatching workload requirements.  For
example, compute cycles may be fully exhausted before memory capacity
is reached, therefore leaving a fraction of the memory unused.
Therefore, the ``disaggregation'' concept has been
proposed where the traditional server's resources
are disaggregated, and then placed in 
shared resource pools~\cite{lim2009disaggregated,lim2012system}.  Servers are
constructed dynamically on-demand by allocating from these resource pools
according to the workload's requirements.  Data gathered from
datacenters show that server memory is unused as much as 50\%
\cite{meng2010efficient,reiss2012heterogeneity,samih2011collaborative,zhang2011characterizing}.
Therefore, it appears at first that disaggregating processors and DRAM
and placing them in their respective pools will be beneficial.
However, we note that a modern microprocessor and DRAM have a very
strong affinity, and in fact they cannot get close enough as evidenced
by the layers of on-chip caches introduced over the years, L1 thru L3
and now an L4 on some processors~\cite{friedrich2014power8}.  Another
issue with memory disaggregation is the signal integrity of long links
required.  Optical interconnects have been proposed as a memory
interconnect~\cite{asghari2011silicon,10.1109/MM.2009.60,tan2009high,chakraborty2012switching,
li1993differential,olsen1993differential,li1993fully,li1991analysis}
as they can provide higher bandwidth over longer distances (e.g., tens
of meters) than electrical interconnects typically with lower power.
Since switching rates and bandwidths of existing memory channels are
couple orders of magnitude higher than the existing electrical fabrics
optical interconnects may be necessary to access remote memory.

In this paper we evaluate the claimed benefits of memory
disaggregation.  We construct a simple cost model that
includes the cost of increased latency, cost of bandwidth and the
datacenter scale savings expected from a disaggregated memory system.
We then identify at which levels a disaggregated memory system is cost
competitive with a conventional direct attached memory system. 
The method can be used to evaluate any future memory technology
that impacts memory cost, latency, or bandwidth.

\section{Disaggregated Memory System}

\begin{figure}
\centering
\resizebox{1.0\linewidth}{!}{\epsfig{file=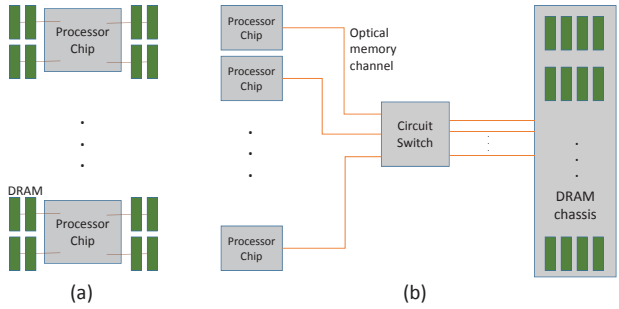}}
\caption{A conventional direct attached memory(a) and a disaggregated memory system(b)}
\label{fig:memoryorganization}
\end{figure}

The conventional direct attached memory organization and a
disaggregated memory organization are shown in
Fig.~\ref{fig:memoryorganization} (a) and (b). In the conventional
memory system, a set of DRAM chips on a circuit board called
dual-inline-memory module (DIMM) typically plugs into to a server
motherboard and connect to the microprocessor typically thru three to
four memory channels implementing DDR3 or DDR4 bus protocols.

On the other hand, the disaggregated memory concept that we evaluate
in this paper has all the DIMMs in a memory pool as shown in
Fig.~\ref{fig:memoryorganization}(b). The memory pool could be a memory
chassis within a server rack, or memory-only racks that serve 
a row of racks, a PoD, or the
entire datacenter.  As such the DIMMs are relatively distant from 
the processors.  

Modern per processor socket memory bandwidths range in the order of 50 to 200
GB/s, about 1 to 2 orders of magnitude higher than 
state of the art network bandwidths.  The processor memory bandwidth depends 
on the number of
memory channels, typically 3-4, and the DRAM frequency. 
To overcome the distance problem in a disaggregated memory system at 
such high data rates optical interconnects may be
necessary in this concept. 
Optical links may carry the
DDR3, DDR4 or yet to be invented future memory channel protocols.  The optical
network may span the entire datacenter or may stay within a rack in a
more modest implementation.  
Processors must implement optical transceivers near by to avoid power
and wiring density issues.  Likewise, memory devices and/or
controllers in the pooled memory must also interface to optical
transceivers.

The memory pool must be connected to the processors through a switch
and memory controllers.  Traditional packet switched networks are
likely to add a prohibitive amount of latency to the memory access
time.  One other option is an Optical Circuit Switch (OCS)
interconnecting all of the pooled processors and pooled memory to
achieve a constant memory latency~\cite{chakraborty2012switching}(an
upcoming workshop of interest~\cite{ofc2015}).  An OCS, for example
the 320 port switch from Calient systems~\cite{calient} contains a
micro-electro-mechanical system (MEMS) of electrically aligned mirrors
that direct light beams from inputs to outputs.  This MEMS based
implementation of OCS has a relatively long circuit setup latency (of
the order 50 milliseconds), however in the disaggregated memory
application it does not matter; the optical circuit will be set up
only once per server configuration.  The pass-through latency through
the switch is negligible as the path from an input port to an output
is merely a couple of mirrors.  With a circuit switch
(optical or electrical) a dedicated path is set up between two end
points: we're assuming in our analysis 
that no queueing or processing delays within the switch occurs.

Using an OCS introduces insertion loss (0.8 to 3.5 dB in the case of
Calient 320x320). Insertion losses may increase bit error rate of the
optical channel and impact performance of the optical transceivers.
It should also be noted that MEMS based optical switches today require
single mode optical transceivers. This is a market in which prices are
coming down relatively quickly due to the maturation of the silicon
photonics technology from more expensive Telco market transceivers to
the relatively low cost data and computer communication transceivers
(which have historically been served by multimode optics).

Note that our cost model neither requires an OCS nor a circuit switch.
A packet switch may also be used at the expense of additional hardware
cost and potential queuing delays in the switch.  The state of the art
memory channel protocols, e.g. DDR3 and DDR4, have strict timing
requirements and can tolerate neither the distance nor any timing
variations in the fabric.  Therefore, we're assuming that a new
memory-channel protocol will be specified by the designers of the
disaggregated memory systems.  Additional memory hierarchies may be
necessary such as L4 or L5 caches.  We discuss caching after
developing the cost model.

Depending on the workload requirements, some number of memory modules
are allocated and circuits are set up between the processor chip and the modules.
We do not envision multiple processor chips splitting the capacity of a memory
module although the cost model does not forbid it.

Note also that the disaggregated memory concept in
Fig.~\ref{fig:memoryorganization} (b) is not a shared memory system;
each memory module is exclusively accessed by a single processor chip
and the memory fabric is not used for interprocessor communication.
Any data sharing and coherency protocols are run across a separate
fabric such as QPI, not shown.

\section{The Cost Model}

The disaggregated memory may introduce orders of magnitude higher
networking bandwidth and additional memory latencies
not present in a direct attached memory system.  In this section, we
construct a simple cost model shown in Eq.~\ref{eq:gain} that
quantifies the cost of latency, cost of bandwidth, and savings to be
achieved from pooling of memory resources. $G$ is the net gain expected
from disaggregated memory, $MS$ is memory savings due to pooling of DRAM resources,
$CL$ is the cost of (increased) latency and $CB$ is the cost of bandwidth:

\begin{equation}\label{eq:gain}
G = MS - (CL + CB)
\end{equation}

Note that $CB$ is the delta cost of optical interconnect (intra or
inter rack) over the on-card electrical link cost of a direct attached
memory which we assumed to be at most \$0.1/Gbps; it includes sockets,
connectors, and a circuit board (and the cost may be even less in low
end systems).  

Note also that Eq.~\ref{eq:gain} is equally applicable to a
``partially disaggregated memory'' where some amount of processor
local memory serves as a cache of the remote disaggregated memory.
Caching reduces the average memory latency which will result in a
smaller cost $CL$ as we will show in the following sections.

\subsection{Cost of latency: performance vs. latency}

\begin{figure}
\centering
\resizebox{1.0\linewidth}{!}{\epsfig{file=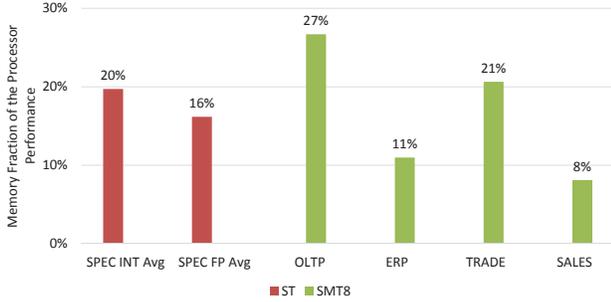}}
\caption{Execution time increase vs memory latency increase summary for SPEC and Commercial benchmarks.}
\label{fig:perfvslat2}
\end{figure}

\begin{figure*}
\centering
\epsfig{file=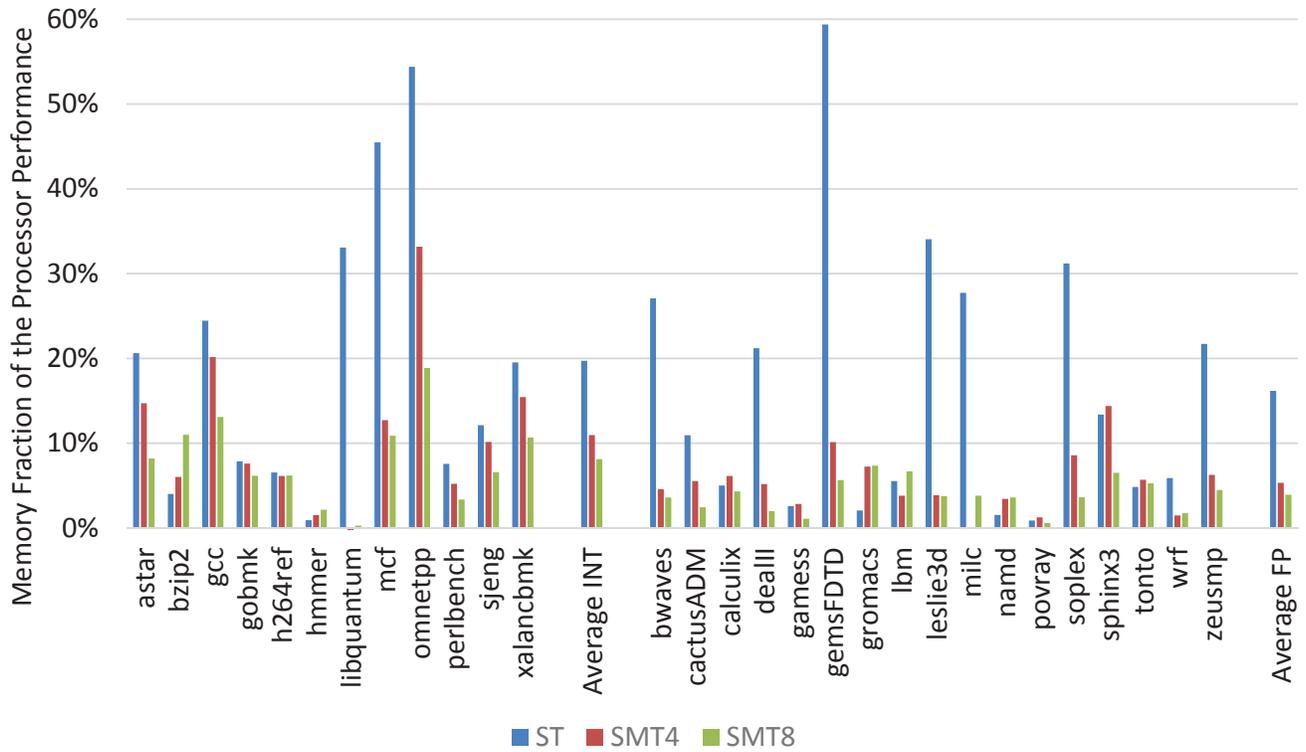}
\caption{Execution time increase as a function of memory latency increase.}
\label{fig:perfvslat}
\end{figure*}

In this section we develop two equations: processor performance as a
function of memory latency and processor price as a function of
performance. By combining the two, we obtain the cost of increased
latency $CL$ due to memory disaggregation.

DRAM latency is an important element of processor performance.  Note
that we are referring to the latency of reading one cache line,
typically 64 to 128 bytes from memory to the processor cache, which
is typically in the range of 50 to 100 nanoseconds.  We're not
referring to the message passing latency over Ethernet or Infiniband
fabrics found in distributed computing systems.  Distributed computing
applications explicitly send and receive messages and therefore they
can amortize and hide the cost messaging latencies by using large
packet sizes and asynchronous messages, which is not practical at the
processor instruction level. If applications are to explicitly access
the remote memory, using RDMA for example, then that scenario would
most likely call for a distributed memory cluster, not the load/store
memory architecture shown in Fig.~\ref{fig:memoryorganization}(b).

We quantified processor performance as a function of memory latency
using a cycle accurate processor simulator.  We simulated 
instruction traces of several benchmarks using different memory
latencies and derived a linear relationship between performance
and latency.  Memory latencies ranged from 75 ns to 300 ns.
Two hypothetical processors P1 and P2 were simulated.  P1
is a 12 core microprocessor with 64KB L1, 512KB L2, and 8MB L3 caches
per core.  P2 is the 16 core version of P1.
Both P1 and P2 use out-of-order execution. 
P1 can issue 10 instructions per cycle to 16 functional
units in each core. Each core can have up to 16 outstanding cache misses.

Both processors implement simultaneous multithreading / hyperthreading found in
x86 and PPC processors~\cite{koufaty2003hyperthreading,sinharoy2011ibm}. 
In the SMT mode each core supports 2, 4 or 8
logical processors that share core's functional units and the execution pipeline,
therefore increasing the total core throughput by a factor of 2 to 3.

The SPEC CPU2006 Integer and Floating Point suite of
benchmark~\cite{speccpu2006} traces were simulated on P2.  The reader
might critique using the SPEC suite here as it is not a ``cloud''
benchmark and it is old etc.  However it is the leading benchmark for
evaluating processor performance which is our focus.  For
example, 15 papers out of 42 used the SPEC suite at the ISCA'2014
symposium.  Four commercial benchmarks OLTP, ERP, TRADE, and SALES
were simulated on P1.  OLTP is an online transaction processing
benchmark that measures the rate of queries/transactions performed on
a database.  TRADE is a Java based stock trading application.  ERP is
an enterprise resource planning application. SALES is a customer order
processing and distribution application.

We quantified workload's sensitivity to memory latency as
the Memory Fraction of Performance ($MFP$) in Eq.~\ref{eq:mfp}

\begin{equation}\label{eq:mfp}
MFP = \Delta ET / \Delta ML
\end{equation}

\noindent where $\Delta ET$ is the percent increase in benchmark 
execution time divided by 
the percent increase in the memory latency, $\Delta ML$, each 
relative to its baseline value.

In essence, MFP is the fraction of execution time attributable to the memory latency.
For example, MFP=40\% indicates that the memory latency is responsible 
for 40\% of the execution time.  Suppose the memory
latency doubled from a base of 75ns to 150ns.  Execution time
would increase by 40\%.  A workload with a small MFP,
e.g. MFP=0\% is insensitive to the memory
latency, because most likely its working set fits in to the processor's
on-chip caches.

Fig.~\ref{fig:perfvslat2} summarizes the memory latency sensitivity
for commercial and SPEC benchmarks.  Fig.~\ref{fig:perfvslat} shows
the memory latency sensitivity of the SPEC 2006 suite of benchmarks in
detail.  Average MFP for the INT and FP suites are 20\% and 16\% on
single threaded (ST) cores.  Individual benchmarks have an MFP as high
as 59\%, which shows the problem that disaggregated memory designers
will face.  For a processor and memory few racks apart
(with a round trip delay of 20m x 5ns/m = 100ns), the total memory 
latency will be more than double the base latency.

A noteworthy observation is that with the increasing SMT levels MFP
decreases for most of the benchmarks.  SPEC INT average MFP=20\% on
the ST core reduces to MFP=8\% on the 8 threaded SMT8 core.  In other
words, workloads are more tolerant to increasing memory latencies on
an SMT processor, consistent with our expectations.  Threads of a core
must wait for each other while accessing the shared functional units
and pipeline stages, which hides some portion of the memory latency.
Results suggest that disaggregated memory systems and other high
latency memory systems may benefit from even higher SMT parallelism.

\begin{figure*}
\centering
\epsfig{file=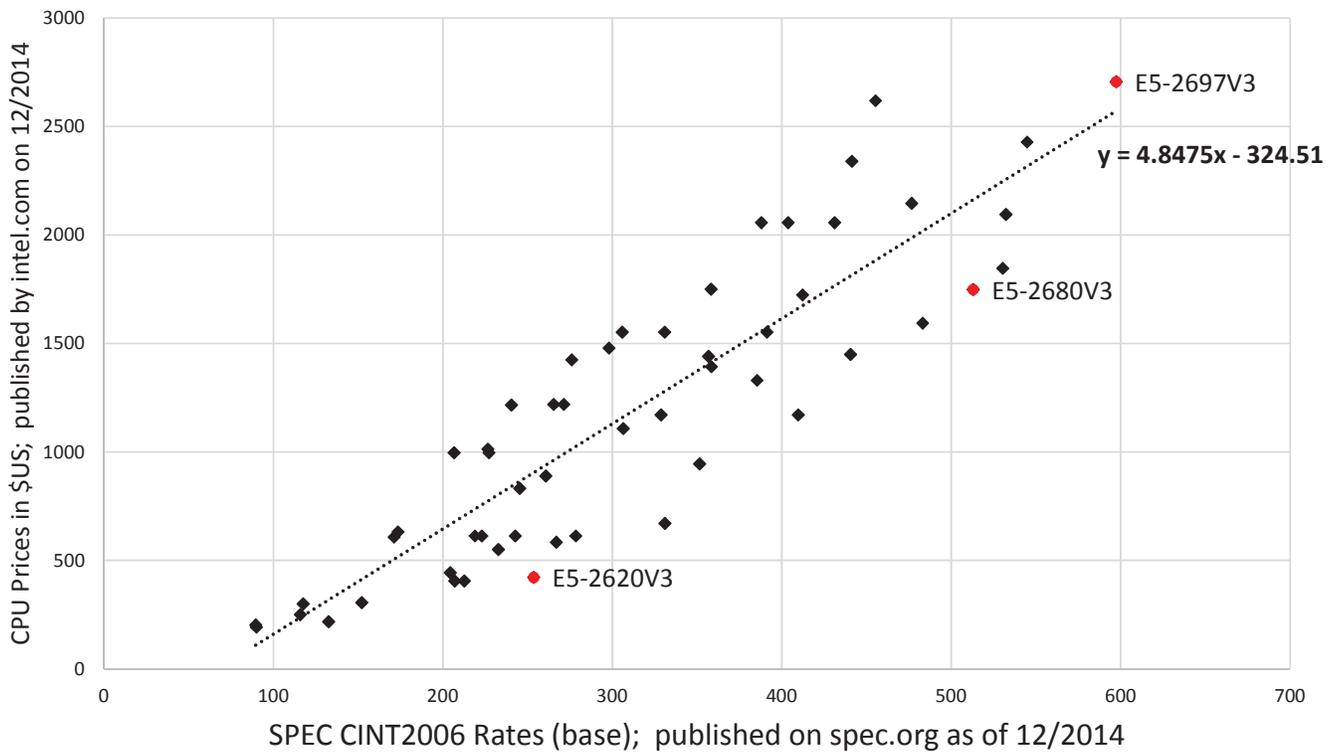}
\caption{Processor price as a function of performance.}
\label{fig:pricevsperf}
\end{figure*}

In a disaggregated memory system, the memory latency may increase
because of distance, queueing/buffering delays if any, electrical/optical
conversion, and protocol delays.  Distance will add about 1 ns to
the latency per 0.2 meter.  A memory chassis in a single rack will add
about 6 meters roundtrip (30 nanoseconds or 40\%) to the base latency of 75ns.  
Memory racks
serving an entire data center will add possibly 100 meters--roundtrip (500ns), or more
to the latency.  Another small source of delay is the serialization and deserialization
(SerDes) function required in a serial optical communication link.
This can be about 10ns, especially if the gear box is a large or an 
odd ratio (e.g. 25/10)~\cite{inphipress}.
Forward Error Correction (FEC) or scrambling is often applied to
the serialized data packets, also adding latency depending
on the FEC code length chosen. 
While these latencies can be negligible today (e.g. 64/66 Ethernet
scrambling or 8b10b block coding), the reduced noise margins at higher data
rates is driving potential use of more complex FECs, e.g. adding 100ns
latency for 4-6dB of link margin~\cite{wang2013}, 
with similar latencies in
electrical backplanes (IEEE 802.3bj clause 91) and adoption into
emerging single mode optics formats~\cite{psm100gspec,clr4spec}.

Some of these variables and design choices are unknown at this time.
Therefore to cover a range of design choices in this paper we used
latency increases from 0\% to 100\% over the base latency of 75ns up
to a max of 150ns.  This range covers a rack scale memory
disaggregation and maybe side by side rack dissagregation.
Processor performance quickly drops with farther distances therefore
we limit our analysis to the 70 -- 150ns range.

\subsection{Cost of latency: price vs performance}

In this section, we quantify the cost of the disaggregated memory
latency.  Suppose the direct attached memory is to be replaced with a
disaggregated memory system.  How would one quantify the monetary
value of the latency increase?   We reasoned that the
datacenter operator would want to compensate or at least know the cost of
any compute cycles
lost to the increased latency.  One approach for compensation may be
deploying more servers in the datacenter.
A cheaper approach may be upgrading to a
higher performance processor along with the disaggregated memory,
therefore having the same datacenter performance before and after.

In our model, we use the higher performance processor's price delta as
a proxy for the cost of increased memory latency. 
(Note that we're not advocating using faster processors with disaggregated
memory systems. We're merely quantifying the cost of latency
via processor price.)  For a given
memory latency increase, we choose a higher throughput processor in
the following manner:

\begin{equation}\label{eq:fastercpu}
Throughput_{newcpu} = \\
(1+ MFP \times Lat_{incr}) \times Throughput_{oldcpu}
\end{equation}

\noindent which will have an equal performance with the old processor 
using a direct attached memory system (where $MFP$ is the workload's
memory fraction of execution time as stated before.)

Note that we're not concerned with the internal organization of this
higher throughput processor.  The performance increase can come from
higher MHz, bigger or additional caches, or more cores.  We need only
the delta cost of the processor for the cost analysis.  Also note that
our method of using processor price as a proxy is general enough that
it may be used to quantify the performance cost of any new memory
technology or memory fabric.

Next, we derive the price and performance relationship:
we retrieved 2240 official benchmark reports from www.spec.org
submitted by computer system vendors for a total of 54 models of Intel
E5-2xxx v2 and v3 series processors.  The SPEC CPU2006 ``base rate'' metric
measures the throughput of a processor. The lowest throughput processor in the 
list, E5-2403v2, has 4 cores, 1.8GHz clock, 10MB cache and has a SPEC integer (INT)
rate of 90.  The highest throughput E5-2697v3, has 14 cores, 2.6GHz clock, 35MB cache
and has an INT rate of 597 (faster E5 processors were on spec.org,
however their prices were not available on intel.com).  Different
vendors reported slightly different rate values for the same model processor,
may be due to the differences in their benchmark configuration or 
software.  We averaged them to arrive at a single rate for the same
model processor.

We then retrieved the prices of E5 v2 and v3 processors from
intel.com (retrieved on Dec. 2014), and plotted the cost vs. performance data point
of each processor model on Fig.~\ref{fig:pricevsperf}.  
We then fitted those data points with a least squares fit
trend-line describing the relationship between cost and performance

\begin{equation}\label{eq:priceperf}
ProcessorPrice = 4.85 \times Throughput - 324
\end{equation}

\noindent Note that the function
is not a perfect fit because price is not necessarily determined only
by the performance.  Processors have other differentiating
characteristics, for example some are low-voltage low-power parts,
some have different QPI bandwidths, some may have been manufactured
with a different process. Intel sets the prices however it chooses.
But, a strong dependence between price and performance is apparent
in Fig.~\ref{fig:pricevsperf}.

\begin{figure}
\centering
\resizebox{1.0\linewidth}{!}{\epsfig{file=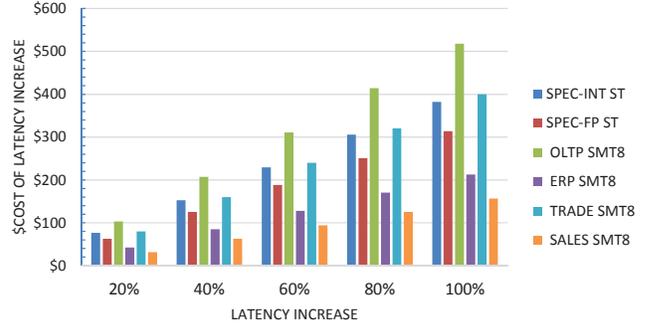}}
\caption{Cost of latency increase}
\label{fig:pricevslat}
\end{figure}

For the sample calculations, we assumed that the conventional direct
attached memory system's processor has a SPEC INT rate of 400.  If for
example memory disaggregation increases execution time by 10\%, then
one could compensate for that loss by choosing a processor with a rate of 440,
and the cost of that +40 rate increase can be estimated from the
fitted line in Fig.~\ref{fig:pricevsperf}.

Combining Eq.~\ref{eq:fastercpu} and Eq.~\ref{eq:priceperf} we arrive at
the cost of increased latency in Fig.~\ref{fig:pricevslat}.  For
example, if the disaggregated memory system increases the latency by
40\% (30 ns), then for SPEC INT and OLTP the cost of that latency
increase are \$155 and \$209, respectively.

\subsection{Cost of bandwidth}

\begin{figure}
\centering
\resizebox{1.0\linewidth}{!}{\epsfig{file=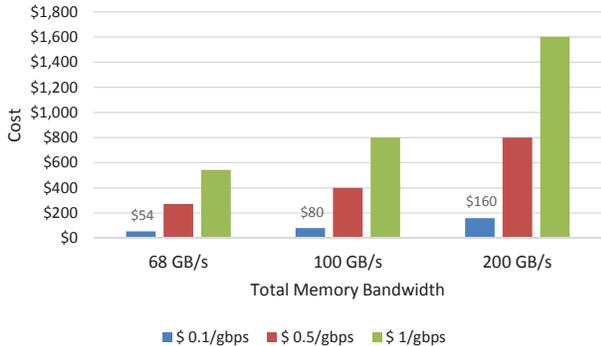}}
\caption{Cost of bandwidth for few unit B/W costs}
\label{fig:costofbw}
\end{figure}

Memory bandwidth requirements of state of the art microprocessors are
a couple orders of magnitude higher than the data communication
network bandwidths.  For example the faster models of Intel E5-2xxx
processors on the right hand side of Fig.~\ref{fig:pricevsperf} all
have a total memory bandwidth of 68 GB/s (GB: gigabytes).  The E7
models, not shown here, have 85 GB/s memory bandwidth.  The IBM Power8
processors have in excess of 230 GB/s of
bandwidth~\cite{friedrich2014power8}.  Therefore, we expect memory
disaggregation to highly burden the networking costs.
Fig.~\ref{fig:costofbw} shows the cost of bandwidth for different unit
costs.  Some historical examples of B/W cost trends for different
optical interconnect form factors~\cite{taub2014} show costs at
several dollars per Gbps in 2012 heading towards \$1/Gbps in 2015. It
should be noted that these are form factor/standard dependent
transceiver costs and that cabling and connector costs can add 25\% or
more (distance dependent).  Furthermore, these transceivers are for
multimode fiber and not suitable for use with OCS MEMS switches
(another component which if used adds additional cost).  The costs for
single mode optical interconnects, however, are declining rapidly with
the maturation of Si Photonics technology, and while more difficult to
predict, could optimistically be assumed to be on par with multimode
transceivers in the not too distant future.

As stated before, there are many designs choices to be made. 
Instead of predicting the future cost of
silicon photonics, we simplify the problem by calculating the unit
bandwidth cost required to make the disaggregated memory system and the
direct attached memory system have an equal-performance and equal-cost.
We set the left hand side of
Eq.~\ref{eq:gain} to zero ($G=0$) and rearrange the terms as

\begin{equation}\label{eq:eqperfeqcost}
CB = MS - CL
\end{equation}

\noindent We can calculate the right hand side of the equation, therefore arriving
at a bandwidth cost $CB$ that has parity with a direct attached memory
system.  As stated before in our calculations we assumed a unit cost of 
\$0.1/Gbps for the direct attached memory interconnect.

\subsection{Disaggregated Memory Savings}

The promise of disaggregated memory is the efficiencies to be achieved
from pooling the memory resources.  Through pooling, unused memory
fragments otherwise locked in individual servers will be made available to
the entire datacenter.  Few data points on unused memory capacities
exist in the literature.  Samih reports that only 69\% of the memory
capacity is used in a cluster of 437 TB size
cluster~\cite{samih2011collaborative}.  Meng reports 45\% reduction in
data center resource requirements when virtual machines with
complementary resource requirements are provisioned
jointly~\cite{meng2010efficient}.  Reiss reports that memory usage
does not exceed about 50\% of the capacity of a Google
cluster~\cite{reiss2012heterogeneity}.  Qi reports memory usage of
five Google backend clusters each of which consisting of thousands of
machines. The five cluster's memory utilitizations are approximately 50, 10, 30,
55, and 65\% respectively~\cite{zhang2011characterizing}.

Using these existing data points, we assume in our example
calculations that memory disaggregation will save as much as 50\% of
the memory capacity.  We also assume that each processor socket has a
maximum 128 GB DRAM attached at a cost of \$7.125 per GB (8 DIMMs x 16
GB/DIMM).  Higher DRAM densities and capacities call for stacked DIMMs
which are very expensive and many cloud providers will not use.  For
example, Softlayer--a cloud provider advertises a maximum of 128GB RAM per socket in its
``bare-metal'' servers.  In sum, 50\% savings of DRAM capacity
translates to \$456 memory savings per processor socket.

\section{Putting things together:  equal cost - equal performance}
\label{sec:equcost}

\begin{figure*}
\centering
\epsfig{file=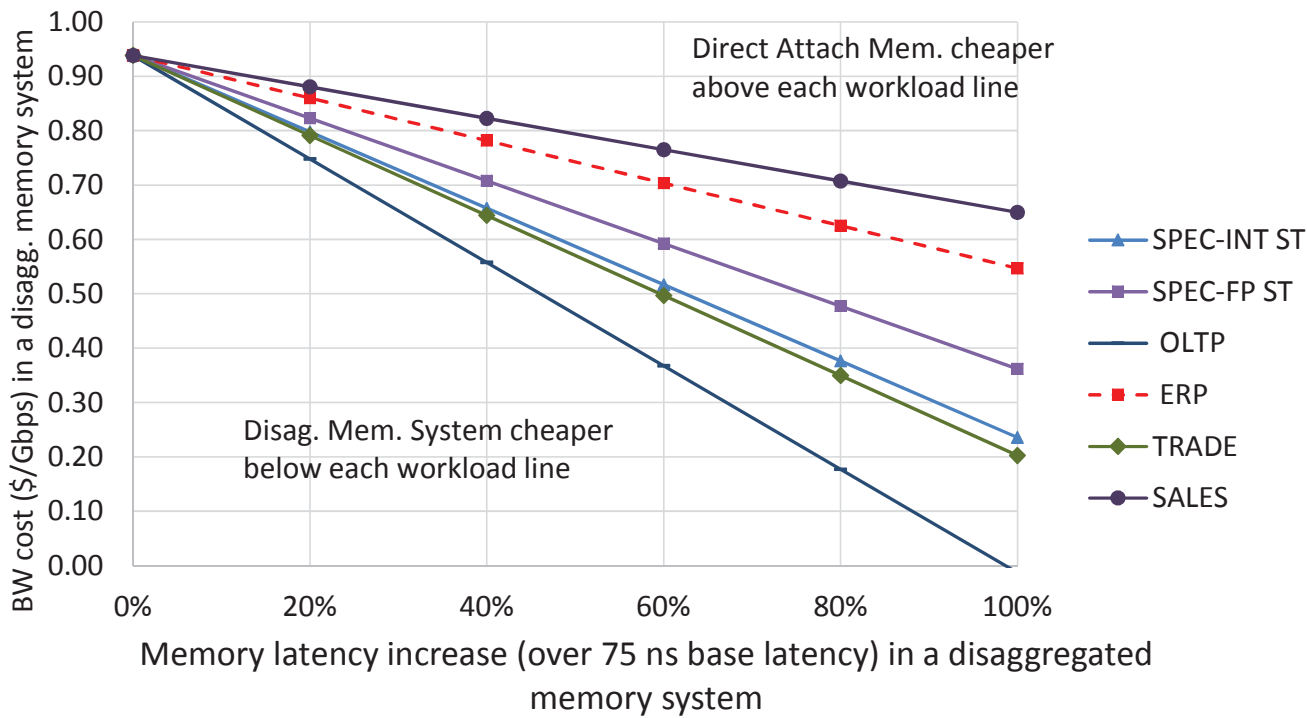}
\caption{Equal-cost equal-performance lines for the disaggregated memory and the direct attached memory.}
\label{fig:equcostequperf}
\end{figure*}

Now that we have quantified both the cost of latency and the savings from
disaggregated memory, we can calculate the unit cost of bandwidth in
terms of dollars per Gbps.

For a given system configuration, workload and memory latency increase 
we lookup the memory savings $MS$ and the cost of latency increase $CL$,
and then calculate $CB$ per Eq.~\ref{eq:eqperfeqcost}. Then
the unit bandwidth cost for an equal cost equal performance system is

\begin{equation}
UnitCost = CB / BW_{memory}
\end{equation}

Note that the unit cost must include all the fabric components shown in
Fig.~\ref{fig:memoryorganization}(b): optical interconnect, the OCS
switch (if used), cabling etc.

Fig.~\ref{fig:equcostequperf} gives the equal cost equal performance
lines for a direct attached memory system and a disaggregated memory
system as a function of the latency increase due to disaggregation.
For a given workload and a latency increase, if the disaggregation
unit cost is below the line, then the disaggregated memory is
cheaper because the memory savings outweigh the cost of
disaggregation.  Otherwise, if the unit cost is above the line, then
the direct attached memory solution is cheaper.  It's apparent from
the set of lines that for a given unit cost --i.e. a horizontal line
across the graph-- as the memory latency increases the savings due to
disaggregated memory diminish.  A rack scale disaggregated memory will
add about 30ns roundtrip for distance alone (40\%). Including the
logic delays a total latency increase of 40\% seems optimistic at the
rack scale. SPEC integer and OLTP workloads require unit costs to be
lower than \$0.70/Gbps and \$0.55/Gbps in this scenario.

It is also apparent that workload performance with the strongest
dependence on memory latency, such as OLTP require unit costs to be
cheaper for disaggregated memory.  Another observation is that the
Y-axis intercept (for a 0\% latency increase) \$0.94/Gbps, is
largely determined by the memory bandwidth requirement of the
processor chip (68GB/s in the example.)  A processor with higher memory
bandwidth will require the disaggregated memory fabric to
be even cheaper and the Y axis intercept must be smaller.  Alternatively, 
to make the case for using a disaggregated memory, the
higher bandwidth processor needs to attach to more than 128GB of direct attached memory to
make the memory savings worthwhile.

\section{Conclusions}

In this paper we developed a simple model of cost vs. performance for
a disaggregated memory system.  The model may also be used for verifying
cost-performance benefits of future memory technologies and memory
fabrics.  

We showed the performance dependency on the memory latency, and that
any latency increase has a non-trivial cost.  We showed that
disaggregated memory needs to have a very low latency to be feasible
and therefore needs to be implemented at the rack level to minimize
the distance.  Additional cache layers may be required in the memory
hierachy to overcome the latency and bandwidth issues.

Our simulations showed that for many workloads, sensitivity
to memory latency decreases with increasing SMT / hyperthread counts.
Therefore, hyperthreaded cores may become an important feature in 
disaggregated systems.

We demonstrated the equal cost and equal performance curves for a
disaggregated memory system. Using today's prices of processor and
memory we arrived at a \$0.94/Gbps cost for
the disaggregated memory fabric with a 0\% latency penalty, and 
approximately \$0.5/Gbps cost with a 50\% latency penalty. 
The unit cost must
include all the component costs from processor's memory channel to
the switch to the DRAM chassis containing the pooled memory.  Actual
cost of photonics fabric are  
in excess of \$1/Gbps for transceiver alone today and possibly \$1.5/Gbps with cabling. 
An OCS switch costs around \$300/port which adds another \$3/Gbps with 100 Gbps links. 
Therefore, the memory dissagregation concept analyzed in this paper
does not appear to be an economical solution.
To justify the practical investment in a disaggregated memory
application, we would like to
see the optical memory interconnect cost to decrease by at least a factor of 3-4 without an OCS and 
a factor of 9-10 with an OCS.
At this time, disaggregation and optical interconnects appear 
more suitable for I/O, for example PCIexpress which is less sensitive
to latency and has lower bandwidth than memory.

Note that recent advances in virtual machines and application
``containers'' may make our 50\% memory savings assumption optimistic,
which is another argument against disaggregated memory. 
Virtualization simplifies bin-packing of workloads in to
underutilized servers.  Furthermore, live migration of VMs and
containers~\cite{isci2011improving,CRIU} can move workloads across the
datacenter transparently and dynamically in response to changing
resource requirements, a property that the statically partitioned
memory pool in Fig.~\ref{fig:memoryorganization}(b) does not have.  As
stated before, Meng reports that by jointly placing VMs with
complementary resources on the same server, datacenter resource
requirements can be reduced by 45\%~\cite{meng2010efficient}.
Isci reports an RDMA based live migration
technique that can migrate virtual machines near the network line rate of 40 Gbps
which will be beneficial if workloads have rapidly changing resource 
requirements~\cite{isci2011improving}.  Therefore, the disaggregated
memory systems of the future may be more relevant to the ``bare-metal''
cloud infrastructures or very large workloads that cannot be colocated
with other workloads.

Our study focused on the acquisition costs. We did not consider the
total cost of ownership, for example potential energy savings due to
the disaggregated memory and silicon photonics over the electrical interconnects.

We did not study the cost-performance of adding another cache layer to the architecture in
Fig.~\ref{fig:memoryorganization}(b).  Han suggests that some amount
of local memory can be a cache of the remote
memory~\cite{Han:2013:NSR:2535771.2535778} and assumes a page level (4KB)
access to the remote memory by exploiting processor's address translation
hardware.
However, we know from our own unpublished work that
software based ``fast paging'' to a remote or slower second tier
memory does not work very well for many workloads. Page fault handling
has a latency of a few microseconds at minimum depending on the
operating system.  Han~\cite{Han:2013:NSR:2535771.2535778} also mentions
this as a caveat. Samih~\cite{samih2011collaborative} shows 
a slowdown of 1 to 2 orders of magnitude in the Hadoop Sort
application when comparing a paged local-to-remote memory system to a system with
enough local memory.  
When a system starts ``fast paging'' to a slower memory
not only the increased latency but the system cycles wasted in the kernel 
reduce workload performance significantly.

Another issue with page level remote memory access is the network
bandwidth explosion.  One hopes for some spatial locality, however
except for sequential access patterns the 4KB size pages over the
fabric often transport large amounts of unused data as the processor's
unit of memory access is only 64 bytes.  

An alternative to the software based paging is implementing a hardware
cache controller and a directory, and using the local DRAM 
as the cache of the
remote DRAM (e.g. a partially disaggregated memory system), and using
a cache block size smaller than a page.  These approaches should perform
better than software based paging, however they have design challenges
and cost to analyze (for
example~\cite{7011373,Qureshi:2012:FLT:2457472.2457502}) that we did
not cover here and we leave it as future work.


%

\bibliographystyle{abbrv}
\bibliography{paper-disag}  


\end{document}